\documentclass[aps,prb,twocolumn,superscriptaddress,reprint,showpacs]{revtex4-1}
\usepackage{graphicx}
\usepackage{overpic}
\usepackage{amsmath}
\usepackage{amssymb}
\usepackage{xcolor}
\usepackage{float}
\usepackage{caption}
\usepackage{subcaption}
\usepackage{tikz}
\usepackage{gensymb}
\usepackage[hypcap=false]{caption} 
\usepackage{silence} 
\definecolor{red} {rgb} {1.0,0.0,0.0}

\begin{document}

\title{Tuned ionic mobility by Ultrafast-laser pulses in Black Silicon}

\author{Christelle Inès K.\ Mebou}
\affiliation{Theoretische Physik, Universit\"at Kassel, Heinrich-Plett-Str. 40, 34132 Kassel, Germany} 
\affiliation{ Center for
  Interdisciplinary Nanostructure Science and Technology 
             (CINSaT), Heinrich-Plett-Str.\ 40, 34132 Kassel, Germany}

\author{Martin E.\ Garcia}
\affiliation{Theoretische Physik, Universit\"at Kassel, Heinrich-Plett-Str. 40, 34132 Kassel, Germany} 
\affiliation{ Center for
  Interdisciplinary Nanostructure Science and Technology 
             (CINSaT), Heinrich-Plett-Str.\ 40, 34132 Kassel, Germany}

\author{Tobias Zier}
\email{tzier2@ucmerced.edu}
\affiliation{Department of Physics, University of California Merced, Merced, CA 95343}

\begin{abstract}
Highly non-equilibrium conditions in femtosecond-laser excited solids cause a variety of 
ultrafast phenomena that are not accessible by thermal conditions, like sub-picosecond solid-to-liquid 
or solid-to-solid phase transitions. In recent years the microscopic pathways of various 
laser-induced crystal rearrangements could be identified and led to novel applications and/or 
improvements in optoelectronics, photonics, and nanotechnology. However, it remains unclear what effect 
a femtosecond-laser excitation has on ionic impurities within an altered crystal environment, 
in particular on the atomic mobility. Here, we performed ab-initio molecular dynamics (AIMD) simulations 
on laser-excited black silicon, a promising material for high-efficient solar cells, using the Code for 
Highly excIted Valence Electron Systems (CHIVES). By computing time-dependent Bragg peak intensities for 
doping densities of 0.16\% and 2.31\% we could identify the overall weakening of the crystal environment 
with increasing impurity density. The analysis of Si-S bond angles and lengths after different excitation 
densities, as well as computing interatomic forces allowed to identify a change in ion mobility with 
increasing impurity density and excitation strength. Our results indicate the importance of impurity 
concentrations for ionic mobility in laser-excited black silicon and could give significant insight 
for semiconductor device optimization and materials science advancement. 
 
\end{abstract}


\maketitle

\section{Introduction}
\label{Intro}

Intense femtosecond-laser excitations induce harsh conditions in solids, mostly because the extremely 
high peak power is mainly transferred to the electronic system, whereas the ions remain nearly unaffected. As 
a consequence, the electronic occupation is changed dramatically, directly influencing the interatomic 
bonding \cite{Graves1998,rethfeld2004timescales,Jiang2018,Gamaly2022}. 
This allows microscopic ionic pathways that are non-accessible under thermal conditions, which are 
the foundation of a variety of ultrafast laser-induced phenomena, like solid-to-liquid phase 
transitions \cite{sokolowski1995ultrafast,Kandyla07,Dhanjal:97,Recoules06,Bengtsson20}, 
solid-to-solid phase transitions \cite{Shumay96,Gaudin12,Giret14,Makita2019,TAVELLA201722,Thomann23}, 
coherent ionic motions \cite{Chang97,fomichev2003laser,zijlstra2013squeezed,Kimata2020,Shayduk22,Ma22,Luo2023},
and even liquid-to-liquid phase transitions\cite{Zalden19}.
Intensive investigations using experimental time-resolved techniques, 
like X-ray \cite{temnov2004ultrafast,Khan20,Pandey2020,Ou2021} or electron diffraction 
\cite{lin2012imaging,Harb08,Yang20,Hada21,HUANG2021} as well as theoretical efforts to simulate 
those effects using 
AIMD \cite{zier2015signatures,Silvestrelli96}, 
Tight-binding approaches\cite{diakhate2010theory,Kim94,Zickfeld99,Medvedev15}, 
classical MD \cite{Ivanov03,SHUGAEV2019,Lipp22,Sakong13}, 
time-dependent DFT \cite{Castro2004,Wachter14,bende2015modeling,Miyamoto2021,Shepard21} 
or other approaches shed light on a variety of underlying microscopic mechanisms and led to fascinating 
applications. In particular, the progress in the field of material processing using ultrafast lasers 
is impressive \cite{malinauskas2016ultrafast, sugioka2017progress}. It's now possible to produce 
structures of the size of critical transport properties \cite{StoianColombier+2020+4665+4688} by 
modifying material surfaces\cite{vorobyev2013direct}, drill nano-precision holes \cite{cheng2017micro}, 
create uniform distributed nano-ripples \cite{hong2014femtosecond}, 
or produce nano-droplets \cite{siew2010investigation} and 
material cones \cite{wang2017direct, tull2006silicon}. 

Besides changing and manipulating crystalline structures by ultrafast-laser light it is possible to alter 
material properties by doping \cite{wang2022aggregation, majid2019review}. Using experimental techniques, 
like ion bombardment \cite{mazey1968observation}, crystal growth \cite{schmidt2009silicon}, sputtering 
\cite{moustakas1979sputtered}, and thermal diffusion \cite{asheghi2002thermal}, it is possible to insert 
different ionic species into a host material. For example, it is an intriguing approach 
\cite{lee2023doped, zhang2021advances, marri2017doped} to introduce dopants to silicon in order to 
manipulate its essential characteristics like the electronic band gap or optical properties. However, 
some methodical challenges appear, like dopant diffusivity \cite{Istratov_2002} or dopant activity 
while maintaining the crystalline structure in the host material \cite{Duffy13}.

Laser-assisted ion migration is combining the benefits of both approaches, in which 
crystal properties can be manipulated twice, 1) by the laser excitation and the ensuing crystal 
changes and 2) by the diffusion of impurities into the crystal. Furthermore, the laser-excitation could 
produce voids or other crystal changes that allow the ions to migrate the crystal in the first place.
In such a way, ions of a gaseous environment could diffuse into the initial crystalline material under 
femtosecond-laser excitation. Among these dopant materials Sulfur (S) has drawn a lot of 
attention  \cite{popelensky2022doping, zhao2014insight, astrov2005}, because of its promising improvements 
for photodiodes, photodetectors, solar cells and antibacterial materials 
\cite{fan2021recent,lv2018review,tan2019nano,otto2015black,liu2014black,casalino2010near,juntunen2016near, 
li2020high,ozkol2020effective, abdullah2016research,liu2019effect, liu2018infinite,lim2018electroosmotic, 
guenther2013investigation, MCKearney23, pirouzfam, mc2023ultrafast}. 
By using silicon as crystal host material in a sulfur environment intense femtosecond-laser excitations 
could produce under certain conditions black silicon \cite{HSU20142,liu2014black,Savin2015}. 
The crystal of black silicon is characterized by the formation of nm to $\mu$m seized needles at the surface 
due to the laser excitation, which incorporate sulfur atoms from the environment. In this configuration, the 
absorption is increased over the range from $250$ to $2500$ nm \cite{Crouch2004} while the reflection is 
reduced to a few percent \cite{Crouch2004,Wang13}. Additionally, experiments show that the photocurrent 
could be increased by up to $60$\% \cite{Sarnet}. We like to note, that black silicon can also be produced 
by other techniques, like dry or wet etching \cite{HSU20142,mullerova2018angle,zhang1989porous}, but the 
use of femtosecond-lasers can reduce the number of fabrication steps \cite{mei2011development}, which in 
general reduces potential influences and/or sources of error. However, the mechanisms behind migration, 
diffusion and distribution mechanisms of defects and/or impurities, here sulfur ions, during and after 
an ultrafast-laser excitation is unknown. 

We performed ab-initio MD simulations of doped 
silicon with an impurity density of $0.46$\% and $2.32$\% and compared it to our reference system of 
pristine silicon. Note, that we don't simulate the needle/cone formation or the capturing of sulfur 
atoms within, rather we studied the mobility change of sulfur dopants in a silicon host crystal. All 
computations were done using the Code for Highly Excited Valence Electron Systems (CHIVES). The rest of 
the paper is structured as followed: Sec. \ref{Simulation_detail} summarizes the main simulation details, 
characteristics of CHIVES, and definitions used for computed quantities. After that, in Sec. \ref{Results} 
we will present our MD simulation results by analyzing, e.g., our obtained time-dependent Bragg peak 
intensities, bond angles and bond lengths of Si-S bonds and interatomic forces on specific atoms. This 
is followed by a discussion of effects contributing to a mobility change and the possibility to tune it 
in Sec. \ref{discussion}.  
  
\section{Simulation details} 
\label{Simulation_detail}

In order to study the mobility change of impurities in laser-excited crystals we compared the structural 
responses of three different supercells with $0$\%, $0.46$\%, and $2.32$\% impurity density. We used 
the data of the supercell with zero impurities as a reference for the silicon atom mobility and refer to it 
in this work as pristine silicon. The supercell itself was constructed by repeating the cubic unit 
cell, which contains $8$ silicon atoms, $3 \times 3 \times 3$ in $x$, $y$, and $z$-direction, respectively. 
Hence, the total number of atoms in the supercell sums up to $216$. In the case of $0.46$\% impurity 
density we replaced randomly one silicon atom of the pristine silicon cell by a sulfur atom. For the 
highest dopant density five randomly chosen silicon atoms were interchanged with sulfur. In order to 
avoid artificial forces or effects we relaxed the doped supercells. For each of the 
three data sets we initialized $10$ independent runs near $315$ K using the approach described in 
\cite{zijlstra2013squeezed}.
We like to note, that in each run five different silicon atoms were interchanged. Moreover, we checked 
that no sulfur clusters were formed, meaning we avoided nearest neighbor atoms to be replaced. The 
lattice parameter for the cubic unit cell is $a = 10.2021 a_0 = 0.5398747278 $ nm. The initialized 
supercells created in this way, were used as input for the MD simulations performed in CHIVES.

\subsection{Specifics used in CHIVES}
\label{Specifics}

CHIVES is a Mermin DFT code that can compute electronic properties for nonzero electronic temperatures 
\cite{zijlstra2013squeezed, zijlstra2013fractional, zier2017simulations}. 
All computations presented here used the local density approximation (LDA) for the 
exchange-correlation functional. The cut-off energy for the Hartree, exchange, and correlation 
potentials was set to 1330 eV. Tightly bonded core electrons are described by norm-conserving 
pseudopotentials \cite{GTH1996}, whereas the valence electrons are accounted for by atom-centered Gaussian 
basis sets. Silicon atoms are described using the following exponents, which were already published in 
\cite{zijlstra2013squeezed}  
$\alpha_1 = 1.77509 a_{0}^{-2}$ ($s$ and $p$ orbitals)
$\alpha_2 = 0.55380 a_{0}^{-2}$ ($s$, $p$ and $d$ orbitals)
$\alpha_3 = 0.16270 a_{0}^{-2}$ ($s$ and $p$ orbitals). 
For the sulfur dopants we established a basis set with $17$ states and the exponents
$\alpha_1 = 1.17769 a_{0}^{-2}$ ($s$ and $p$ orbitals)
$\alpha_2 = 0.40348 a_{0}^{-2}$ ($s$, $p$ and $d$ orbitals)
$\alpha_3 = 0.12989 a_{0}^{-2}$ ($s$ and $p$ orbitals).
The perturbation of the electronic system by a femtosecond-laser excitation is modelled by an 
instant increase of the electronic temperature. That corresponds to a $\delta$-pulse excitation above 
the materials band gap and the approximation that the electrons thermalize fast, namely within the first 
time step. Consistent to that electronic temperature the electrons are distributed over the Kohn-Sham 
states using the corresponding Fermi-Dirac distribution. Interatomic forces computed by CHIVES are then 
incorporated into a Velocity-Verlet scheme to connect electrons and ions and to enable MD simulations. 
Here, we used a timestep of $\Delta t = 1$ fs. Highly parallelized subroutines allow to perform 
propitious calculations, in particular on non-symmetric and/or disordered systems.

\subsection{Bragg peak intensities}
\label{Bragg_Intensities}

With the aim of gaining information about the structural deformations and variations of black silicon 
after femtosecond-laser irradiation we computed the time-evolution of Bragg peak intensities, which 
could directly be compared to data obtained by time-resolved diffraction experiments using X-rays 
or electrons. In general, the intensity $I$ can be expressed in terms of the time-dependent structure 
factor $F$:  

\begin{equation}
    I_{\textbf{q}}(t) = \left| F_{\textbf{q}}(t) \right |^2  
\end{equation}

with $\textbf{q} = \frac{2\pi}{a} (h,k,l)$ as the scattering vector and $h$, $k$, $l$ the Miller indices defining 
the scattering plane. Given that our crystal contains two different types of atoms, Si and S, the 
structure factor itself can be expressed as the total of the contributions made by each species:

\begin{equation}
  F_{\textbf{q}}(t) =  \sum_{j=1}^{N_1} f_{j,1} e^{ i\textbf{q}.r_{j,1}(t)} +  \sum_{k=1}^{N_2} f_{k,2} e^{ i\textbf{q}.r_{k,2}(t)}
\end{equation}

\begin{figure*}[ht!]
\centering
  \begin{subfigure}{.33\textwidth}
    \includegraphics[width=1\linewidth]{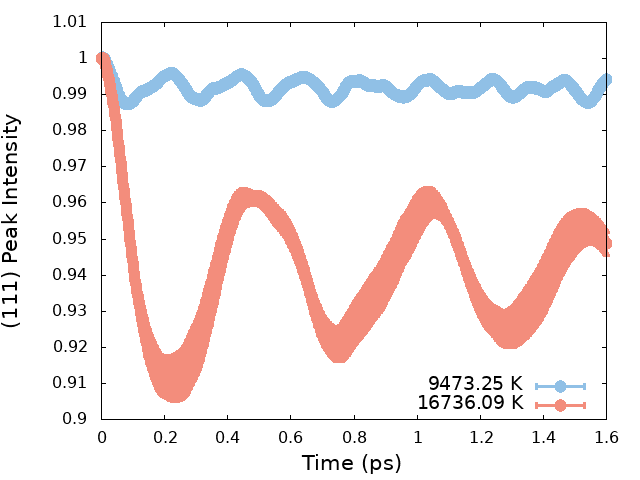}
    \caption{}
  \end{subfigure}%
  \begin{subfigure}{.33\textwidth}
    \includegraphics[width=1\linewidth]{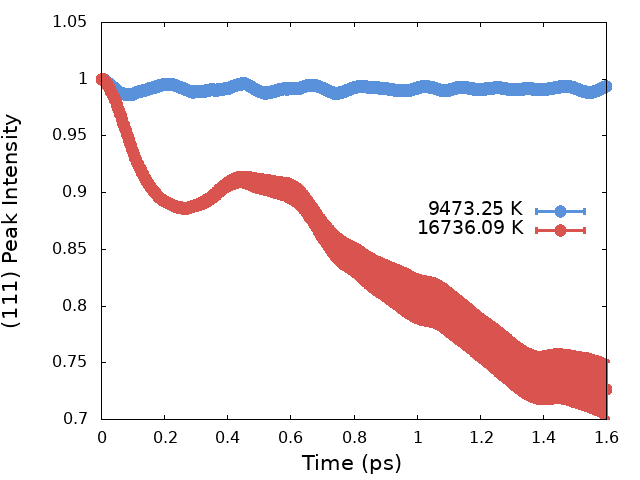}
    \caption{}
  \end{subfigure}%
  \begin{subfigure}{.33\textwidth}
    \includegraphics[width=1\linewidth]{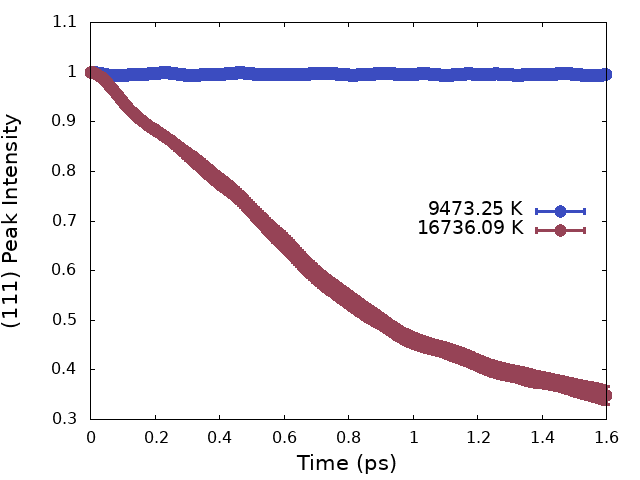}
    \caption{}
  \end{subfigure}%
\caption{Normalized time-evolution of the (111) Bragg intensity after femtosecond-laser excitations that 
induce electronic temperatures $T_1 = 9473.25$ K (blue-ish solid curves) and $T_2 =16736.09$ K (red-ish 
solid curves), respectively. (a) Laser-induced coherent oscillations in pristine silicon. (b) For a 
dopant density of $0.46$\% the crystal destabilizes for the higher excitation with $T_2$. (c) This trend is 
even accelerated for the dopant density of $2.31$\%. The crystal remains stable for all densities and an 
excitation to $T_1$. Note, the different scale in intensity between the densities. The width of the lines 
indicate the error in the average.}
\label{Bragg_111}
\end{figure*}

with $N_1$ and $N_2$ the total number of atoms of each type in the supercell, 
$f_{j,1} = 14$ 
and $f_{k,2} = 16$ 
the scattering factors of the $j-th$ atom of type 1 (Si) and the $k-th$ atom of type 2 (S)\cite{prince2004international}, $r_{j,1}$ and $r_{k,2}$ 
the position vectors of the $j-th$ atom of type 1 and the $k-th$ atom of type 2 in the unit cell, 
respectively. We note, that the intensity in the presented form is dependent on the number of atoms in 
the supercell. With the intention of getting rid of this dependency we normalized 
$I_{\textbf{q}}(t=0)$ to $1$. Note, that the intensities shown in Fig.\ \ref{Bragg_111} are averaged 
over our ten independent runs. The width is indicating the errors in the averages.

\subsection{Si-S bond lengths and angles}
\label{Bonding}

Another insight of the ionic dynamics after the laser-excitation is provided by the evolution of the 
Si-S bond length. We obtained an approximation of the bond length $b$ by averaging the distance to first 
neighbor silicon atoms from each sulfur dopant using

\begin{equation}
b = \frac{1}{N_{S} N_{nn}} \sum_{i=1}^{N_{S}} \sum_{j=1}^{N_{nn}} |\textbf{r}_{i}^{S} - r_{j}^{Si}| \;,
    \label{eq:length}
\end{equation}
with $N_{S}$ is the number of dopants, $N_{nn}$ is the number of nearest neighbors, $\textbf{r}_{S}$ is 
the position of a sulfur dopant, and $r_{j}^{Si}$ is the position the dopants nearest Si neighbors.
Again, we averaged this quantity over our ten independent runs. \\

Moreover, we computed the Si-S-Si bond angle $\theta$ and its time-evolution by using
\begin{equation}
\theta = \frac{1}{6 N_{S}} \sum_{i=1}^{N_{S}} \sum_{i,j > i}^{6} arccos \left( \frac{\textbf{r}^{Si-S}_{i}\cdot\textbf{r}^{Si-S}_{j}}{|\textbf{r}^{Si-S}_{i}||\textbf{r}^{Si-S}_{j}|}
 \right) \; ,
    \label{eq:angle}
\end{equation}
with $\textbf{r}^{Si-S}_{i,j} = \textbf{r}_{i,j}^{Si} - \textbf{r}^{S}$. 


\section{AIMD Simulation results} 
\label{Results}

Coming back to our initial question ''how is the mobility of dopants changed within a host crystal after 
an ultrashort optical excitation?" we analyzed the above described simulations. In the following, we show 
and compare the results for two different excitation strength, realized by two electronic temperatures, 
namely $T_1 = 30$ mHa $\approx 9473.25$ K and $T_2 = 53$ mHa $\approx 16736.09$ K. $T_1$ corresponds to a 
moderate excitation for which no irreversible structural changes are induced \cite{Zier2014}. $T_2$ is 
close to but below the threshold of laser-induced disordering processes that will definitely destroy the 
crystal symmetry.   
 
In a first attempt, we analyzed the evolution of Bragg peak intensities after the excitation (see
Fig.\ \ref{Bragg_111}). In the left panel of Fig.\ \ref{Bragg_111}(a) the results for pristine silicon are 
plotted. Without any impurities present in the crystalline system the intensity only decreases by around 
$9$\%, even for the higher temperature. This indicates that no irreversible or disordering processes are 
induced within the simulation timescale. For both electronic temperatures oscillations in the Bragg 
intensities are visible. Those can be attributed to thermal phonon squeezing \cite{zijlstra2013squeezed}.
Surprisingly, an impurity density of $0.46$\%, here a single Si atom was interchanged by S, has already a 
drastic impact on the overall system's response to optical excitations (see Fig.\ \ref{Bragg_111} (b)). 
Whereas the moderate excitation to $T_1$ seems comparable to pristine silicon, the case for $T_2$ came out 
completely different. Within the simulation time the intensity drops by roughly $30$\%, indicating 
disordering processes in which atoms move far away from their equilibrium positions and destroying 
the crystal symmetries. Interestingly, the first stages, up to $400$ fs, show a similar behavior as 
pristine silicon, namely an oscillatory behavior (see Fig.\ \ref{Bragg_111} (b), red curve). However, 
the oscillation does not continue and the intensity drops.
\begin{figure*}
\centering
  \begin{subfigure}{.5\textwidth}
    \centering
    \includegraphics[width=1\textwidth]{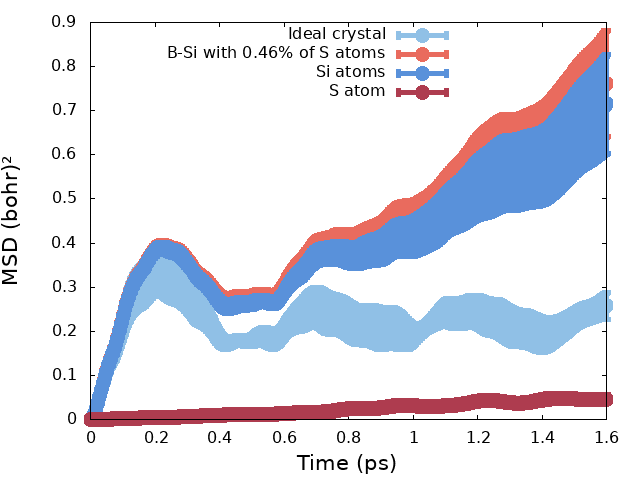} 
  \end{subfigure}%
  \begin{subfigure}{.5\textwidth}
    \centering
    \includegraphics[width=1\textwidth]{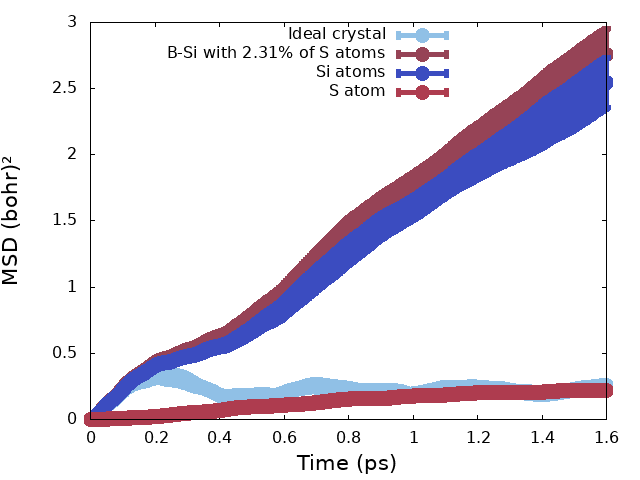} 
  \end{subfigure}%
\caption{Averaged MSD for an electronic temperature $T_2 = 16736.09$ K as a function of time and a dopant 
density of $0.46$\% (left panel, light red) and $2.31$\% (right panel, red-brown), respectively. The 
values for pristine silicon is given as a reference (light blue) in both figures. The decomposition into 
the silicon host crystal (blue) and the sulfur impurities (red, bottom curve).}
\label{msd53}
\end{figure*}
Thereafter the atoms follow other pathways by 
overcoming laser-changed local bonding, which accelerates them away from their initial local minimum.     
In addition we like to point out that within the first $400$ fs the width of the curve in 
Fig.\ \ref{Bragg_111} (b) is relatively small, which means that basically all atoms in all independent runs 
behave at least very similar if not the same. After this period the width increases, indicating different 
atomic behavior between the runs, which is typical at transition thresholds. In summary, the interchange of 
a single silicon atom with sulfur reduced the threshold for irreversible structural changes and increased 
the ionic mobility within the structure. This trend is even aggravated after increasing the dopant density 
to $2.31$\% (see Fig.\ \ref{Bragg_111} (c)). The intensity drops by almost $70$\% within the simulation 
time. Moreover, there are no longer any signs of oscillations visible, the transient maximum is only 
slightly visible at around $200$ fs. The dopant density has therefore a direct influence on the crystal's 
response after femtosecond-laser excitation and the ion mobility. Note, that the results for an additional 
Bragg direction can be found in Appendix. However, this analysis of the Bragg 
intensity is an average over the whole crystal, which makes it impossible to decide which amount can be 
attributed to the movement of sulfur atoms and what comes from the silicon host material. Therefore, we 
computed the mean-square atomic displacement (MSD) by

\begin{align} \label{msd}
MSD(t,0)  & =  \frac{1}{N_1 + N_2}  \sum_{i=1}^{N_1} {|r_{i}(t)-r_{i}(0)|}^2 \notag\\
    & + \frac{1}{N_1 + N_2}  \sum_{j=1}^{N_2} {|r_{j}(t)-r_{j}(0)|}^2 \;.
\end{align}
Here, $N_1$ and $N_2$ are the numbers of atoms for each type, $r_{i,j}(t)$ is the position vector of 
atom $i$,$j$ at time $t$, $r_{i,j}(0)$ the initial position vector of atom $i$,$j$ at time $t=0$. 
Figure \ref{msd53} shows the results for $T_2 = 16736.09$ K for both dopant densities $0.46$\% (left panel) 
and $2.31$\% (right panel). Compared to pristine silicon (light blue), the MSD of $0.46$\% shows an 
increasing behavior (light red in Fig.\ \ref{msd53}) to around $0.75$ bohr$^2$. This is in accordance to 
the intensity drop in the (111) Bragg peak we discussed before. Furthermore, Fig.\ \ref{msd53} shows the 
decomposition of the total MSD into the contributions from the silicon host crystal (blue in Fig.\ \ref{msd53}) 
and the sulfur impurities (dark red Fig.\ \ref{msd53}). We observed that most is contributed to the total MSD 
by the silicon host crystal. However, we recognized a monotonic increased contribution by the sulfur atoms. 
A similar behavior, but on a different scale could be observed in the case of the higher dopant density. 
In this case the sulfur atoms reach the average displacement of pristine silicon within the simulation time.   
Surprisingly, this result indicates an enormous mobility increase in the host material and not for the 
impurities. This means, in the case of mobility that the host crystal benefits of the impurities, or in 
other words that the impurities destabilize the crystalline structure without moving exorbitantly themselves.
\begin{figure}[h!]
     \centering
         \includegraphics[width=0.5\textwidth]{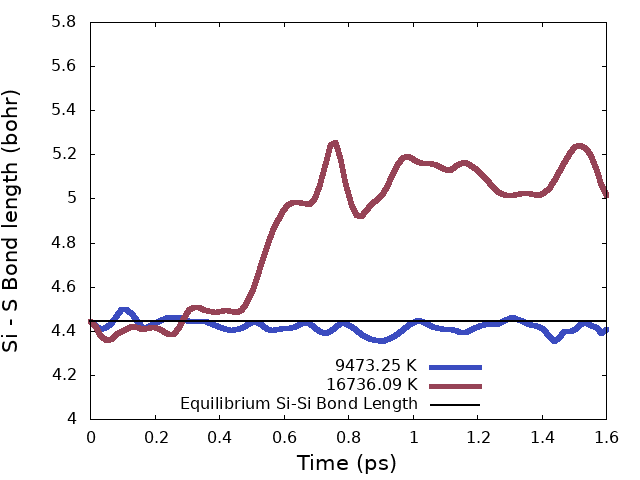}
         \caption{Bond length evolution of impurities to neighboring atoms as a function of time after the 
         excitation for a dopant density of $2.31$\%. }
         \label{length}
\end{figure}

In order to investigate the changes occurring in the vicinity of the impurities we changed to quantities 
that are sensitive to microscopic changes, like the bond length of impurity atoms to nearest neighbors. 
Since our results indicated a mostly similar crystal response, except the scale, we will from here on only 
show results for the high dopant density of $2.31$\%. The results for the lower density can be found in 
Appendix. Figure \ref{length} shows the bond length evolution after the excitation. For the 
lower excitation to $T_1$ the bond length stays close to its initial value with a small decreasing trend at 
the end of the simulation, which indicates that the microscopic arrangement around the impurity is remaining 
diamond-like and the atoms don't move that far. For the higher excitation (red curve in Fig.\ \ref{length}) 
the bond length increased about $10$\% within our simulation. This is also observable by the shift to larger 
values of the nearest neighbor distance in the pair-correlation function (see Appendix) at this electronic 
temperature. Interestingly, the bond length change is not a monotonic increase but rather showing some dips. 
It starts with a shortening of the interatomic bond length within the first $200$ fs, which could be 
attributed to the induced pressure by the excitation. After that the bond length increases by almost $10$\% 
before showing a big dip at around $800$ fs. At around this time the MSD starts to increase linearly with 
time, indicating that the system behaves diffusive. Therefore, scattering events become more probable that 
could change the direction of motion of the impurities or the surrounding ions, which could also push the 
impurity back towards its initial position. Nevertheless, on average the distance of impurities to its 
neighbors increase. 

\begin{figure}[h!]
     \centering
         \includegraphics[width=0.5\textwidth]{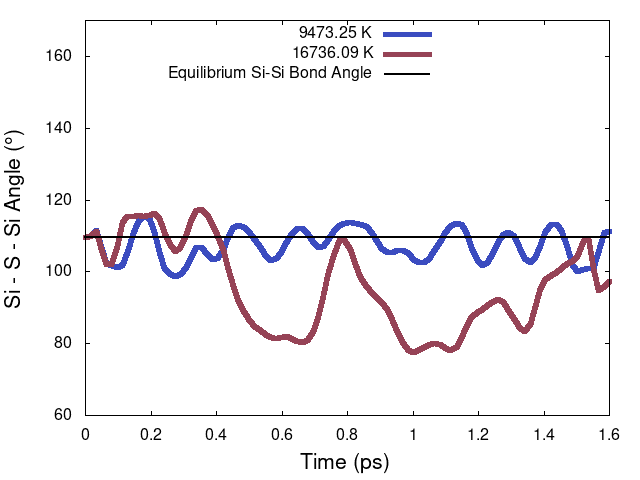}
         \caption{Time-evolution of the averaged bond angle in the vicinity of an impurity.}
         \label{angle}
\end{figure}

We see a similar behavior in the time-evolution of the bond angle 
in the vicinity of sulfur atoms (See Fig.\ \ref{angle}). The results for an excitation to $T_1$ (red line 
in Fig.\ \ref{angle}) shows an oscillation around the the ideal angle of $109.5$\degree, which means that 
the diamond-like structure is maintained. The interatomic bonding remains mostly intact, the ions oscillate 
around their initial positions without destroying so much the crystal's symmetry.   
In the case of a higher excitation, here $T_2$ (Fig.\ \ref{angle}, blue curve), we can also see some 
oscillations of the bond angle, in particular in the first $400$ fs. After that period it drops to around 
$80$\degree, a change of almost $30$\%, indicating a dramatic change on the microscopic environment around 
the sulfur dopants. Again, we see a rebound to the initial angle at around $800$ fs, a possible recovery 
of the tetrahedral, diamond-like structure. Then we see a recurring drop in the bond angle with a 
subsequent slow increase. 

\begin{figure*}
\centering
  \begin{subfigure}{.5\textwidth}
    \centering
    \includegraphics[width=1.1\textwidth]{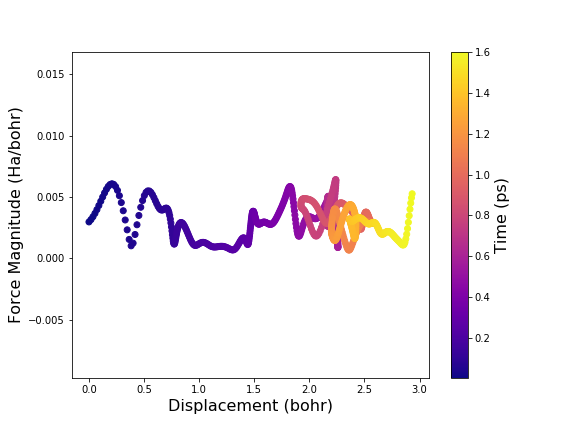} 
  \end{subfigure}%
  \begin{subfigure}{.5\textwidth}
    \centering
    \includegraphics[width=1.1\textwidth]{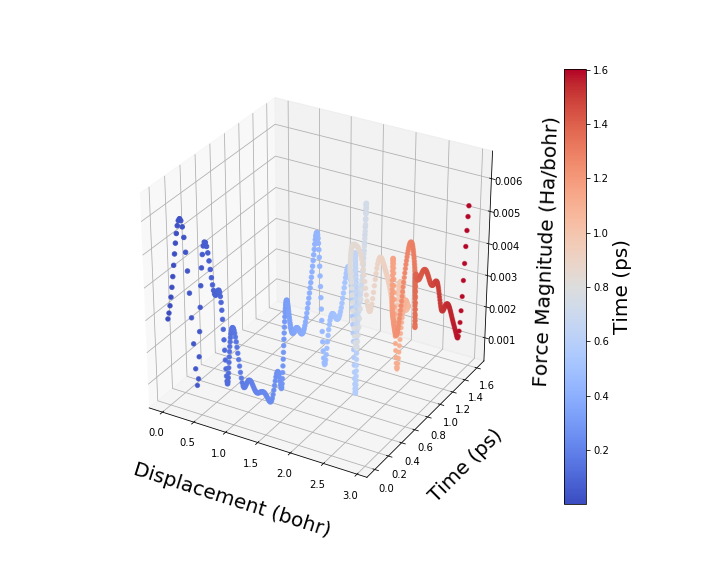} 
  \end{subfigure}%
\caption{(left panel) Interatomic forces acting on sulfur atoms as a function of their root-mean-square 
atomic displacement. The color indicates the time after the excitation, when the corresponding displacement 
is reached. (right panel) 3D plot of the interatomic forces as a function of the displacement and time after 
the excitation. In this way the knot-like structure was unraveled.}
\label{Forces}
\end{figure*}

In order to shed light on this behavior we analyzed the interatomic forces acting 
on the sulfur atoms as a function of displacement. CHIVES computes the interatomic forces for every timestep 
of the simulation on the fly, meaning that the potential energy surface (PES) that describes the interatomic 
bonding is updated every timestep to the new positions of the ions. It's gradient gives the interatomic forces. 
The averaged absolute value of the force is shown in Fig.\ \ref{Forces} (left panel). In the first $600$ fs, 
indicated by the color code, we see a constant increase in the atomic displacement, while the forces 
during this time, with exception of the initial acceleration during the first $150$ fs, do not show 
large variations. In other words, the impurities are accelerated after the excitation due the laser-changed 
bonding in the crystal, but move relatively free through the crystal. After about $700$ fs the ions reach 
a root-mean-square displacement between $2.0$ and $2.5$ bohr and remain in this range for another $700$ fs. 
Several direction changes and peaks in the force magnitude indicate a larger number of scattering events during 
this time of the simulation. The 3D plot of the forces (see Fig.\ \ref{Forces}, right panel) as a function 
of the time after the laser-excitation and the root-mean-square displacement unravel the knot structure 
between $2.0$ and $2.5$ bohr in the 2D plot. 
By comparing those results to the ones with lower sulfur density (see Appendix) at the same level of excitation 
we recognized that in the latter case the sulfur ions on average move farther away from their initial position. 
In addition, the scattering that causes substantial direction changes takes place at much later times. As a 
result, the ionic mobility of the dopants depend strongly on the laser-excitation but also on the dopant 
density. A higher dopant density will have a larger impact on the stability of the crystal. As a consequence 
the ions of the host material will start to move more easily, which will cause ion scattering much earlier 
than at lower dopant densities. In other words, a higher dopant density will improve the host materials 
mobility, which will decrease the dopant mobility due to much more often occurring scattering events. 

\begin{figure}[ht!]
    \centering
    \includegraphics[width=0.5\textwidth]{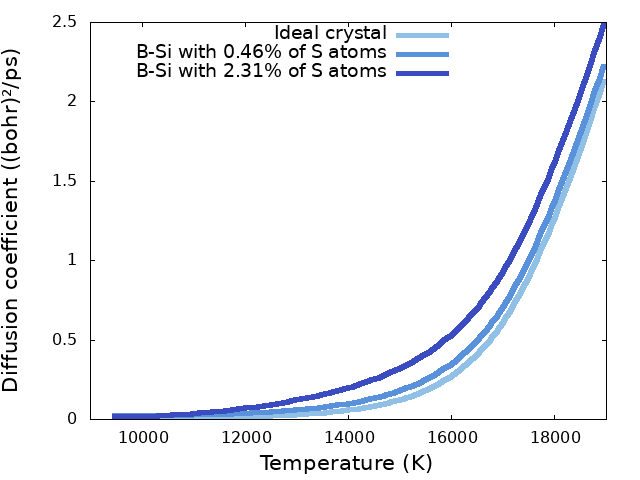} 
    \begin{tikzpicture}[overlay, remember picture]
        \coordinate (inset) at (-2.5,2.5);
        \node[draw, inner sep=0pt, anchor=south west] at (inset) {\includegraphics[width=0.23\textwidth]{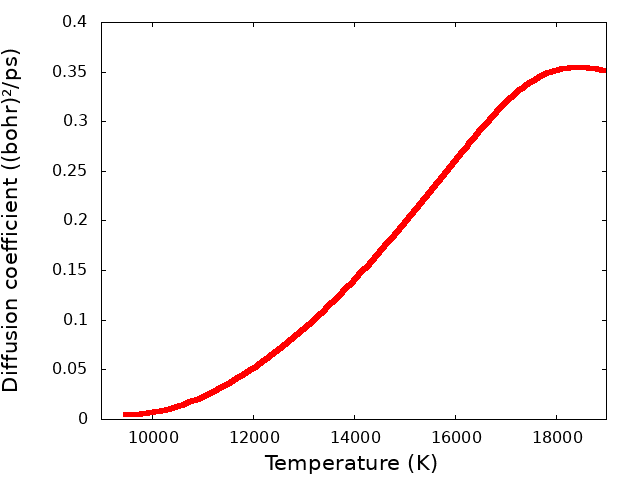}}; 
    \end{tikzpicture}
    \caption{Diffusion coefficient as a function of electronic temperature for three different dopant densities $0$, $0.46$\%, and $2.31$\%, respectively. The inset shows the difference between the diffusion coefficients of our supercell with $2.31$\% of S atoms and the pristine silicon.}
    \label{diff}
\end{figure}

Our observation that the impurities have a strong impact on the host crystal and its ion mobility can 
also be seen in the diffusion coefficient. It is a measure of the rate at which particles move within a 
material as a result of temperature fluctuations or outside disturbances. Figure \ref{diff} shows the 
electronic temperature dependence of the diffusion coefficient for pristine silicon (light blue), silicon 
with one sulfur impurity per supercell ($0.46$\%, blue), and silicon with sulfur density of $2.31$\% (dark 
blue). The diffusion coefficient $D$ was computed from MSD (See Eq. \ref{msd}) by using the Einstein relation:
\begin{equation}
    D =\frac{1}{6}\lim_{t \rightarrow \infty} \left( \frac{d}{dt} MSD (t)\right)
\end{equation}
In agreement with previous observation in this work we see at each electronic temperatures that the diffusion 
coefficient is largest for the highest dopant density ($2.31$\%), showing the highest mobility but also 
the most scattering events. The inset in Fig.\ \ref{diff} shows the difference between pristine silicon and 
silicon with the highest dopant density in this work. We observe that for increasing electronic temperature 
the difference becomes larger before it reaches a plateau or even decrease again. The reason for that is, 
at some excitation level the crystalline structure destabilized predominantly due to the extremely laser-
changed bonding properties, which become much larger than the locally changed properties due to impurities.

\section{Effects on the impurity mobility} 
\label{discussion}

In summary, our present work on sulfur-doped silicon has shed light on the dynamic behaviour after a 
femtosecond-laser excitation on the microscopic level. In more detail, by performing ab initio molecular 
dynamics, we studied the motion and mobility of sulfur atoms in a silicon host crystal. Furthermore, we  
analyzed the ionic mobility in dependence of the dopant density and the laser-excitation strength. 
Our results show, that at the same level of excitation, here electronic temperature, the increase of 
dopants has a direct influence on the threshold after which irreversible structural changes are induced by 
the excitation. The time-evolution of Bragg peak intensities indicates that this threshold is reduced with 
increasing number of dopants. The weakening of the crystal's bonding can be explained by the additional 
change of the local bonding by the dopant. Interestingly, this has a larger effect on the ions of the host 
material than on the dopants. The contributions of the host crystal and the dopants to the total MSD show 
that on average each silicon ion moved twice as far away than in the pristine comparison system (Fig.\ 
\ref{msd53}, left panel) or even more, whereas the dopant contribution is relatively small. 
Analyzing the 
dopant vicinity by calculating bond angles and lengths, we found that the diamond like structure is nearly 
preserved for moderate excitation but can undergo drastic changes at higher excitation strengths. For example 
can the bond angle change from the equilibrium value of $109.5$\degree to below $80$\degree. However, our 
results indicate several rebounds, at least in the bond angle, which implies that the dopant also stabilizes 
its crystal vicinity. The bond length in this case changes to larger values as also seen in the shift of the 
nearest neighbor peak in the pair-correlation function. This causes more scattering events and a slow down 
of ions in general. In the case of high dopant density and electronic temperature the ions are more likely 
to get stuck at earlier times and smaller displacements due to ion-ion scattering that at lower densities or 
pristine crystals. Generating black silicon by altering the crystalline structure by femtosecond-laser pulses 
and the migration of sulfur atoms into the material from the environment can now be tuned by two parameters, 
the excitation strength and the dopant density. Our results show, that the dopant mobility can be tuned over 
a relatively large region but going to the extremes is not beneficial. In future works, we like to investigate 
the dopant mobility by additionally including vacancies in the system. Furthermore, we are interested in 
following the microscopic pathways of the migration process. In silicon all lattice sites are defined by 
symmetry, therefore, it would be of interest to study the tuning of ion mobility in materials with more 
degrees of freedom.

\begin{acknowledgments}
Calculations for this research were conducted on the Lichtenberg high performance computer of the 
TU Darmstadt and the IT Service center (ITS) University of Kassel. C.I.K.M. was supported by the 
Doctoral scholarship from the University of Kassel. T.Z. was supported by the Deutsche 
Forschungsgemeinschaft (DFG) through the projects GA 465/18-1 and ZI 1858/1-1. M.E.G. acknowledges 
support from the DFG through project GA 465/27-1.
\end{acknowledgments}

\appendix
\section{Additional data}
\label{Additional_data}

In order to have a concise description of the main points of our work and improve the readability we 
moved some data used in this work to the Appendix. The interested reader can find in the following the 
pair-correlation function, additional Bragg peak intensities and the interatomic forces for the low density dopant supercell. 

\subsection{Pair-correlation-function}

\begin{figure*}
\centering
  \begin{subfigure}{.5\textwidth}
    \centering
    \includegraphics[width=1.1\textwidth]{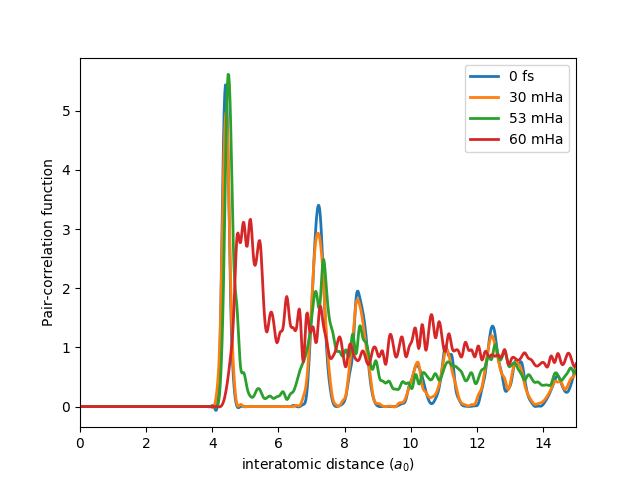} 
  \end{subfigure}%
  \begin{subfigure}{.5\textwidth}
    \centering
    \includegraphics[width=1.1\textwidth]{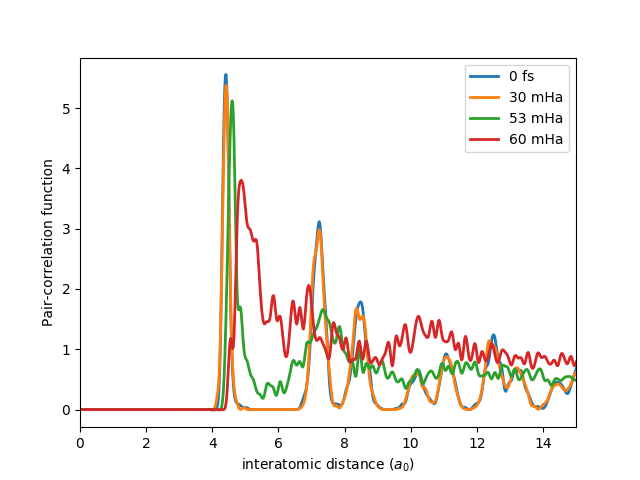} 
  \end{subfigure}%
\caption{Pair-correlation function for a dopant density of $0.46$\% (left panel) and $2.31$\% (right panel), 
respectively, for different electronic temperatures. As a reference we plotted the ground state distribution 
at $t=0$ fs (blue). All other distributions are computed for the ionic distribution at the end of our 
simulation.}
\label{pcf}
\end{figure*}

In a highly symmetric crystal accumulations of atoms are found at particular distances, which are 
characteristic for the given material. The pair-correlation function (pcf) is a quantity that scans the 
crystal and counts all atoms at a certain distance. It is possible to obtain the pcf by 
\begin{equation}
g(r) = \sum_{i,j} \frac{G(r-r_{ij})}{2 \pi r^2 N} \; ,
\label{eq:direct_pcf}
\end{equation}
with $G$ a Gaussian with full width at half maximum of $0.1$ {\AA} and $r_{ij}$ the distance between atom 
$i$ and $j$. The Gaussian is used to smoothen the results of our finite supercell. Figure \ref{pcf} 
(left panel) shows the results of the pcf for a dopant density of $0.46$\%. As a reference we plotted the 
ground state distribution at $t=0$ fs (blue curve). The three other curves correspont to different electronic 
temperatures, namely $T_1 = 30$ mHa $\approx 9473.25$ K (orange), $T_2 = 53$ mHa $\approx 16736.09$ K (green), 
and additionally $T_3 = 60$ mHa $\approx 18946.52$ K (red). All pcf shown here are computed for ionic 
configuration corresponding to the last timestep in our simulation. $t=1.6$ ps. For $T_1$ only small changes 
in the magnitude can be recognized even for large interatomic distances. Note, that the positions of the peaks 
remain the same, indicating the existing of the diamond-like crystalline structure with all symmetries. 
Increasing the electronic temperature to $T_2$ introduces noticeable changes to the distribution. First, the 
long range order got washed out by merging peaks. Second, increasing peaks between the main peaks of the ground 
state. Third, the nearest neighbor peak at around $4.4$ bohr shifts slightly to larger distances. Increasing 
the electronic temperature ($T_3$) even more causes the total loss of crystalline symmetry. No distinguished 
peaks are recognizeable, the function transitions to the form of a liquid system. In the case of higher 
dopant density (see Fig.\ \ref{pcf}, right panel) we see a similar behavior for $T_1$. The pcf in the case 
of $T_2$ shows also a loss of long range order but a more pronounced movement of the nearest neighbor peak 
to longer distances. This is in agreement with our finding of the increased bond length after laser-excitation. 

\begin{figure*} 
\centering
  \begin{subfigure}{.33\textwidth}
    \includegraphics[width=1\linewidth]{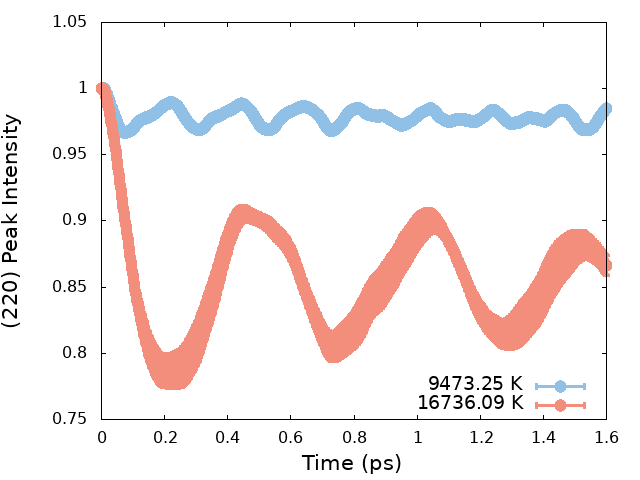}
    \caption{}
  \end{subfigure}%
  \begin{subfigure}{.33\textwidth}
    \includegraphics[width=1\linewidth]{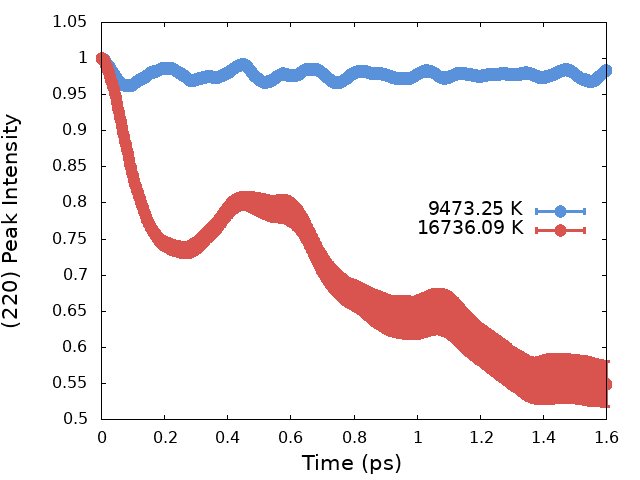}
    \caption{}
  \end{subfigure}%
  \begin{subfigure}{.33\textwidth}
    \includegraphics[width=1\linewidth]{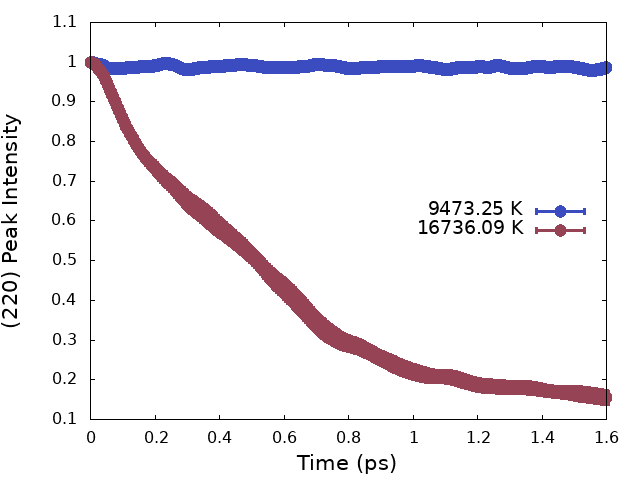}
    \caption{}
  \end{subfigure}%
\caption{Normalized time-evolution of the (220) Bragg intensity after femtosecond-laser excitations that 
induce electronic temperatures $T_1 = 9473.25$ K (blue-ish solid curves) and $T_2 =16736.09$ K (red-ish 
solid curves), respectively. (a) Laser-induced coherent oscillations in pristine silicon. (b) For a 
dopant density of $0.46$\% the crystal destabilizes for the higher excitation with $T_2$. (c) This trend is 
even accelerated for the dopant density of $2.31$\%. The crystal remains stable for all densities and an 
excitation to $T_1$. Note, the different scale in intensity between the densities. The width of the lines 
indicate the error in the average.}
\label{Bragg_220}
\end{figure*}
\subsection{(220) Bragg peak intensity}

In addition to the time-evolution of the $(111)$ Bragg peak intensities we added the intensities for the  
$(220)$ peak here. Figure \ref{Bragg_220} shows the intensities for the $(220)$ direction at the same 
conditions mentioned in the main text. In fact, the same behavior can be observed only the scale of the 
effect is slightly larger.

\subsection{Interatomic forces on impurities}

\begin{figure*}
\centering
\subfloat[]{\includegraphics[width=0.5\textwidth]{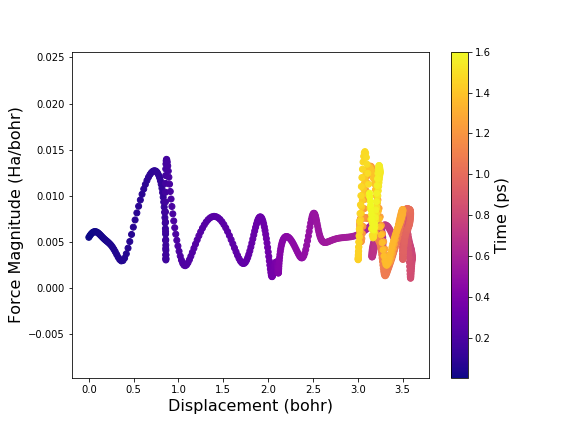}}\hspace{-0.8cm}
\subfloat[]{\includegraphics[width=0.5\textwidth]{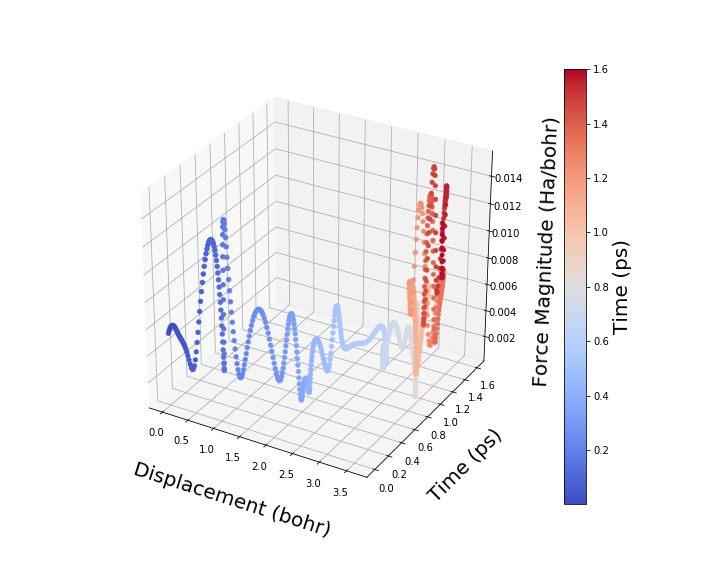}}

\vspace{0cm}
\subfloat[]{\includegraphics[width=0.5\textwidth]{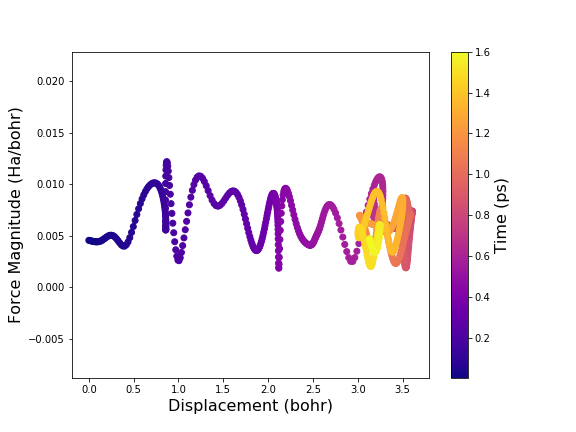}}\hspace{-0.8 cm}
\subfloat[]{\includegraphics[width=0.5\textwidth]{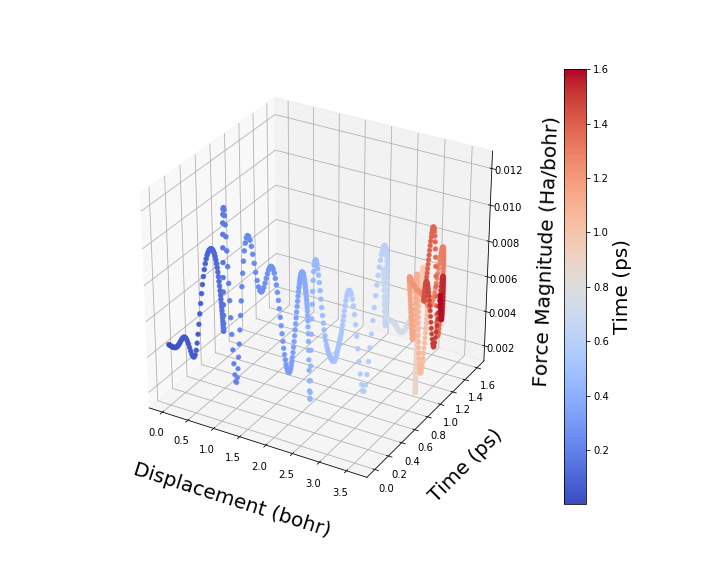}}
\vspace{0cm}
\subfloat[]{\includegraphics[width=0.5\textwidth]{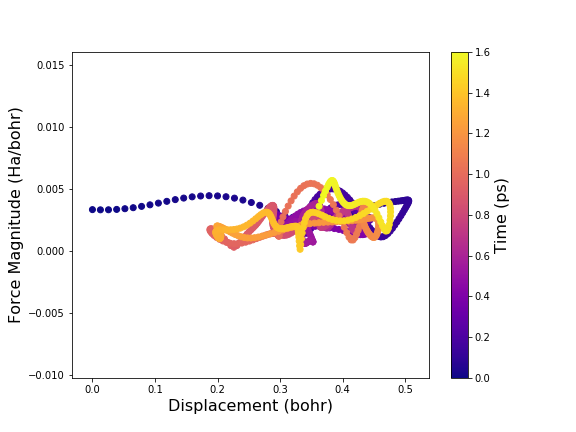}}\hspace{-0.8 cm}
\subfloat[]{\includegraphics[width=0.5\textwidth]{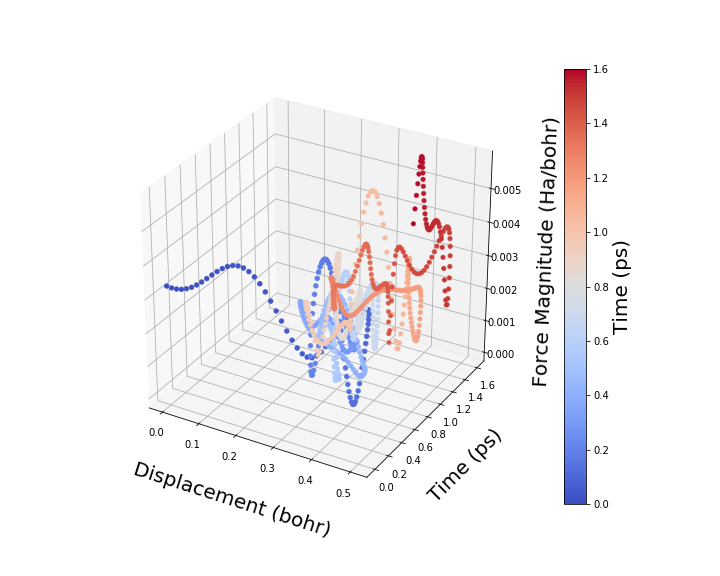}}
\caption{Forces acting on Sulfur atoms as a function of atoms displacements (left panels) and as a function of 
atomic displacement and time (right panels) for different electronic temperatures and dopant densities.} 
\label{fu53}
\end{figure*}

For completeness we show here the interatomic forces data for the low dopant density of $0.46$\% 
and electronic temperature $T_1$ (Fig.\ \ref{fu53}, (a) and (b)) and $T_2$ (Fig.\ \ref{fu53}, (c) and (d)).
In addition, we show the results for the high dopant density $2.31$\% and electronic temperature $T_1$ (Fig.\ \ref{fu53}, (e) and (f)).

\newpage
\bibliography{Literature}

\end{document}